\documentclass[prl,aps,twocolumn,showpacs]{revtex4}
\usepackage{times,epsfig,amssymb,amsfonts,amsmath}
\newcommand{\ket}[1]{|#1\rangle}
\newcommand{\bra}[1]{\langle #1|}
\newcommand{\proj}[1]{\ket{#1}\bra{#1}}

\begin{document}

\title{General Monogamy Relation for the Entanglement of Formation in Multiqubit Systems}

\author{Yan-Kui Bai$^{1,2}$}
\email{ykbai@semi.ac.cn}
\author{Yuan-Fei Xu$^1$}
\author{Z. D. Wang$^2$}
\email{zwang@hku.hk}

\affiliation{$^1$ College of Physical Science and Information
Engineering and Hebei Advance Thin Films Laboratory, Hebei Normal
University, Shijiazhuang, Hebei 050024, China\\
$^2$ Department of Physics and Centre of Theoretical and Computational
Physics, The University of Hong Kong, Pokfulam Road, Hong Kong, China}

\begin{abstract}
We prove exactly that the squared entanglement of formation, which quantifies the bipartite
entanglement, obeys a general monogamy inequality in an arbitrary multiqubit mixed state.
Based on this kind of exotic monogamy relation, we are able to construct two sets of useful
entanglement indicator: the first one can detect all genuine multiqubit entangled states
even in the case of the two-qubit concurrence and $n$-tangles being zero, while the second
one can be calculated via quantum discord and applied to multipartite entanglement dynamics.
Moreover, we give a computable and nontrivial lower bound for multiqubit entanglement of
formation.
\end{abstract}


\maketitle

For multipartite quantum systems, one of the most important properties is that entanglement is
monogamous \cite{horo09rmp}, which implies that a quantum system entangled with another system
limits its entanglement with the remaining others~\cite{ben96pra}. For entanglement quantified by the squared
concurrence \cite{wootters98prl}, Coffman, Kundu, and Wootters (CKW) proved the first quantitative
relation \cite{coffman00pra} for three-qubit states, and Osborne and Verstraete proved the corresponding
relation for $N$-qubit systems, which reads~\cite{osborne06prl}
\begin{equation}\label{1}
C_{A_1|A_2\cdots A_N}^2-C_{A_1A_2}^2-C_{A_1A_3}^2\cdots-C_{A_1A_N}^2\geq0
\end{equation}
Similar inequalities were also generalized to Gaussian systems~\cite{adesso06prl,hiroshima07prl}
and squashed entanglement \cite{koashi04pra,christandl04jmp}. As is known, the monogamy
property can be used for characterizing the entanglement structure in many-body
systems~\cite{coffman00pra,byw0708pra}. A genuine three-qubit entanglement measure named
``three-tangle" was obtained via the monogamy relation of squared concurrence in three-qubit pure states
\cite{coffman00pra}. However, for three-qubit mixed states, there exists a special kind of
entangled state that has neither two-qubit concurrence nor three-tangle
~\cite{lohmayer06prl}. There also exists a similar case for $N$-qubit mixed states~\cite{bai08pra2}.
To reveal this critical entanglement structure other exotic monogamy relations beyond the squared
concurrence may be needed.

On the other hand, from a practical viewpoint, to calculate the entanglement measures appeared in the
monogamy relation is basic. Unfortunately, except for the two-qubit case~\cite{wootters98prl}, this task
is extremely hard (or almost impossible) for mixed states due to the convex roof extension of pure state
entanglement~\cite{ben96pra2}. Quantum correlation beyond entanglement (e.g., the
quantum discord \cite{ollivier01prl,vedral01jpa}) has recently attracted considerable attention, and
various efforts have been made to connect quantum discord to quantum entanglement~\cite{modi12rmp}.
It is natural to ask whether or not the calculation method for quantum discord can be
utilized to characterize the entanglement structure and entanglement distribution in multipartite
systems.

In this Letter, by analyzing the entanglement distribution in multiqubit systems, we prove exactly
that the squared entanglement of formation~\cite{wootters98prl} is monogamous in an arbitrary
multiqubit mixed state. Furthermore, based on the exotic monogamy relation, we construct two sets
of useful indicators overcoming the flaws of concurrence, where the first one can detect all genuine
multiqubit entangled states and be utilized in the case when the concurrence and $n$-tangles are zero,
while the second one can be calculated via quantum discord and applied to a practical dynamical
procedure. Finally, we give a computable and nontrivial lower bound for multiqubit entanglement
of formation.

\emph{General monogamy inequality for squared entanglement of formation.} -- The entanglement of
formation in a bipartite mixed state $\varrho_{AB}$ is defined as \cite{ben96pra2,wootters97prl},
\begin{equation}\label{2}
E_f(\varrho_{AB})=\mbox{min}\sum_i p_i E_f(\ket{\psi^i}_{AB}),
\end{equation}
where the minimum runs over all the pure state decompositions $\{p_i,\ket{\psi^i}_{AB}\}$ and
$E_f(\ket{\psi^i}_{AB})=S(\rho_A^i)$ is the von Neumann entropy of subsystem $A$. For a two-qubit
mixed state $\rho_{AB}$, Wootters derived an analytical formula \cite{wootters98prl}
\begin{equation}\label{3}
E_f(\rho_{AB})=h(\frac{1+\sqrt{1-C_{AB}^2}}{2}),
\end{equation}
where $h(x)=-x\mbox{log}_2x-(1-x)\mbox{log}_2(1-x)$ is the binary entropy and
 $C_{AB}=\mbox{max}\{0,\sqrt{\lambda_1}-\sqrt{\lambda_2}-\sqrt{\lambda_3}-\sqrt{\lambda_4}\}$ is
the concurrence with the decreasing nonnegative $\lambda_i$s being the eigenvalues of the matrix
$\rho_{AB}(\sigma_y\otimes \sigma_y)\rho_{AB}^*(\sigma_y\otimes \sigma_y)$.

A key result of this work is to show exactly that the bipartite entanglement quantified by the
squared entanglement of formation $E_f^2$ obeys a general monogamy inequality in an arbitrary
$N$-qubit mixed state, \emph{i.e.},
\begin{equation}\label{4}
E_f^2(\rho_{A_1|A_2\cdots A_n})-E_f^2(\rho_{A_1A_2})-\cdots -E_f^2(\rho_{A_1A_n})\geq 0,
\end{equation}
where $E_f^2(\rho_{A_1|A_2\cdots A_n})$ quantifies the entanglement in the partition
$A_1|A_2\cdots A_n$ (hereafter $n=N$ for qubit cases), and $E_f^2(\rho_{A_1A_j})$ quantifies the
one in the two-qubit system $A_1A_j$.
Under two assumptions, a qualitative analysis on three-qubit pure states was given in Ref.
\cite{bai13pra}. Before showing the general inequality, we first give the two propositions,
whose analytical proofs are presented in the Supplemental Material~\cite{Suppl}.

\emph{Proposition I}:  The squared entanglement of formation $E_f^2(C^2)$ in two-qubit mixed
states varies monotonically as a function of the squared concurrence $C^2$.

\emph{Proposition II}:  The squared entanglement of
formation $E_f^2(C^2)$ is convex as a function of the squared concurrence $C^2$.

We now analyze the monogamy property of $E_f^2$ in an $N$-qubit pure state
$\ket{\psi}_{A_1A_2\cdots A_n}$. According to the Schmidt decomposition \cite{peres95book},
the subsystem $A_2A_3\cdots A_n$ is equal to a logic qubit $A_{2\cdots n}$. Thus the entanglement
$E_f(A_1|A_{2}\cdots A_n)$ can be evaluated using Eq. (3), leading to
\begin{eqnarray}\label{5}
&&E_f^2(C^2_{A_1|A_2\cdots A_n})\nonumber\\
&\geq& E_f^2(C_{A_1A_2}^2+\cdots +C_{A_1A_n}^2)\nonumber\\
&\geq& E_f^2(C_{A_1A_2}^2)+E_f^2(C_{A_1A_3}^2)+\cdots+E_f^2(C_{A_1A_n}^2),
\end{eqnarray}
where we have used the two propositions, with the details presented in \cite{Suppl}.

At this stage, most importantly, we prove that the squared entanglement of formation $E_f^2$ is
monogamous in an arbitrary
$N$-qubit mixed state $\rho_{A_1A_2\cdots A_n}$. In this case, the analytical Wootters formula in
Eq. (3) cannot be applied to $E_f(\rho_{A_1|A_2\cdots A_n})$, since the subsystem $A_2A_3\cdots A_n$
is not a logic qubit in general. But, we can still use the convex roof extension of pure state
entanglement as shown in Eq. (2). Therefore, we have
\begin{eqnarray}\label{6}
E_f(\rho_{A_1|A_2\cdots A_n})=\mbox{min}\sum_i p_i E_f(\ket{\psi^i}_{A_1|A_2\cdots A_n}),
\end{eqnarray}
where the minimum runs over all the pure state decompositions $\{p_i,\ket{\psi^i}\}$. We assume
that the optimal decomposition for Eq. (6) takes the form
\begin{equation}\label{7}
\rho_{A_1A_2\cdots A_n}=\sum_{i=1}^{m} p_i \ket{\psi^i}_{A_1A_2\cdots A_n}\bra{\psi^i}.
\end{equation}
Under this decomposition, we have
\begin{eqnarray}\label{8}
&&E_f(\rho_{A_1|A_2\cdots A_n})=\sum_i p_i E_f(\ket{\psi^i}_{A_1|A_2\cdots A_n})=\sum_i E1_i\nonumber\\
&&E_f^\prime(\rho_{A_1A_j})=\sum_i p_i E_f(\rho^i_{A_1A_j})=\sum_i Ej_i,
\end{eqnarray}
where the $E_f^\prime(\rho_{A_1A_j})$ is the average entanglement of formation under the specific
decomposition in Eq. (7) and the parameter $j\in[2,n]$. Then we can derive the following monogamy
inequality
\begin{eqnarray}\label{9}
&&E^2_f(\rho_{A_1|A_2\cdots A_n})-\sum_j E_f^{\prime 2}(\rho_{A_1A_j})\nonumber\\
&=&(\sum_i E1_i)^2-\sum_j(\sum_iEj_i)^2\nonumber\\
&=&\sum_i(E1_i^2-\sum_j Ej_i^2)\nonumber\\&&+2\sum_i\sum_{k=i+1}(E1_iE1_k-\sum_jEj_iEj_k)\geq 0,
\end{eqnarray}
where, in the second equation, the first term is non-negative because the $E_f^2$ is monogamous in
pure state components, and the second term is also non-negative from a rigorous analysis shown in
the Supplemental Material~\cite{Suppl},
justifying the monogamous relation. On the other hand, for the
two-qubit entanglement of formation, the following relation is satisfied
\begin{equation}\label{10}
E_f(\rho_{A_1A_j})\leq E_f^{\prime}(\rho_{A_1A_j}),
\end{equation}
since the $E_f^{\prime}(\rho_{A_1A_j})$ is a specific average entanglement under the decomposition
in Eq. (7), which is greater than $E_f(\rho_{A_1A_j})$ in general. Combining Eqs. (9) and (10),
we can derive the monogamy inequality of Eq. (4), such that we have completed the whole
proof showing that the squared entanglement $E_f^2$ is monogamous in $N$-qubit mixed states.

\emph{Two kinds of multipartite entanglement indicator.} -- Lohmayer \emph{et al} \cite{lohmayer06prl}
studied a kind of mixed three-qubit states composed of a $GHZ$ state and a $W$ state
\begin{equation}\label{11}
\rho_{ABC}=p\proj{GHZ_3}+(1-p)\proj{W_3},
\end{equation}
where $\ket{GHZ_3}=(\ket{000}+\ket{111})/\sqrt{2}$, $\ket{W_3}=(\ket{100}+\ket{010}+\ket{001})/\sqrt{3}$,
and the parameter $p$ ranges in $[0,1]$. They found that, when the parameter $p\in (p_c,p_0)$ with
$p_c\simeq0.292$ and $p_0\simeq 0.627$,
the mixed state $\rho_{ABC}$ is entangled but without two-qubit concurrence and three-tangle.
The three-tangle quantifies the genuine tripartite entanglement and is defined as \cite{coffman00pra}
$\tau(\rho_{ABC})=\mbox{min}\sum_i p_i [C^2_{A|BC}(\ket{\psi^i_{ABC}})-
C^2_{AB}(\rho^i_{AB})-C^2_{AC}(\rho^i_{AC})]$.
It is still an unsolved problem on how to characterize the entanglement structure in this kind of
states, although an explanation via the enlarged purification system was given~\cite{bai08pra2}.

Based on the monogamy inequality of $E_f^2$ in pure states, we can introduce a kind of indicator for
multipartite entanglement in an $N$-qubit mixed state $\rho_{A_1A_2\cdots A_n}$ as
\begin{equation}\label{12}
\tau_{SEF}^{(1)}(\rho^{A_1}_N)=\mbox{min}\sum_i p_i[E_f^2(\ket{\psi^i}_{A_1|A_2\cdots A_n})
-\sum_{j\neq 1}E_f^2(\rho^i_{A_1A_j})],
\end{equation}
where the minimum runs over all the pure state decompositions $\{p_i,\ket{\psi^i}_{A_1A_2\cdots A_n}\}$.
This indicator can detect the genuine three-qubit entanglement in the mixed state specified
in Eq. (11). After some analysis, we can get the optimal pure state
decomposition for the three-qubit mixed state
\begin{equation}\label{13}
\rho_{ABC}=\frac{\alpha}{3}\sum_{j=0}^{2}\proj{\psi^j(p_0)}+(1-\alpha)\proj{W_3},
\end{equation}
where the pure state component $\ket{\psi^j(p_0)}=\sqrt{p_0}\ket{GHZ_3}-e^{(2\pi i/3)j}
\sqrt{1-p_0}\ket{W_3}$ and the parameter $\alpha=p/p_0$ with $p<p_0\simeq 0.627$.
Then the indicator is
\begin{eqnarray}\label{14}
\tau_{SEF}^{(1)}(\rho_{ABC}^A)&=&\alpha\tau_{SEF}^{(1)}(\ket{\psi^0(p_0)})
+(1-\alpha)\tau_{SEF}^{(1)}(\ket{W})\nonumber\\
&=&\alpha\cdot s_p+(1-\alpha)\cdot s_w
\end{eqnarray}
where $s_p\simeq0.217061$ and $s_w\simeq0.238162$. In Fig.1, we plot the entanglement indicators
$\tau_{SEF}^{(1)}$, $E_f^2(AB)+E_f^2(AC)$ and $E_{f}^2(A|BC)$ in comparison to the indicators $\tau$,
$C_{AB}^2+C_{AC}^2$ and $C^2_{A|BC}$ calculated originally in Ref.~\cite{lohmayer06prl}.
As seen from Fig.1, although the three-tangle $\tau$ is zero when $p\in(0,p_0)$, the nonzero
$\tau_{SEF}^{(1)}$ indicates the existence of the genuine three-qubit entanglement.
This point may also be understood as a fact that the three-tangle $\tau$ indicates merely
the $GHZ$-type entanglement while the newly introduced indicator $\tau_{SEF}^{(1)}$ can
detect \emph{all} genuine three-qubit entangled states.

\begin{figure}
\begin{center}
\epsfig{figure=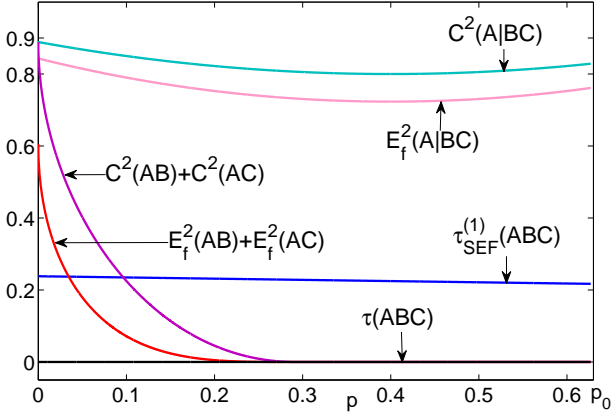,width=0.45\textwidth}
\end{center}
\caption{(Color online) Entanglement indicators $\tau_{SEF}^{(1)}$, $E_f^2(AB)+E_f^2(AC)$ and
$E_{f}^2(A|BC))$ in comparison with the indicators $\tau$, $C_{AB}^2+C_{AC}^2$ and $C^2_{A|BC}$
in Ref. \cite{lohmayer06prl}, where the nonzero $\tau_{SEF}^{(1)}$ detects the genuine
three-qubit entanglement in the region.}
\end{figure}

For three-qubit mixed states, a state $\varrho_{ABC}$ is called genuine tripartite-entangled if 
any decomposition into pure states $\varrho_{ABC}=\sum_i p_i \proj{\psi^i_{ABC}}$ contains at least one
genuine tripartite-entangled component $\ket{\psi_{ABC}^i}\neq \ket{\phi_1}\otimes\ket{\phi_2}$
with $\ket{\phi_1}$ and $\ket{\phi_2}$ corresponding to the states of a single qubit or a couple of qubits
\cite{horo09rmp}. For the tripartite entanglement indicator $\tau_{SEF}^{(1)}(\varrho_{ABC}^A)$, we
have the following lemma and the proof can be found in the Supplemental Material \cite{Suppl}.

\emph{Lemma 1}: For three-qubit mixed states, the multipartite entanglement indicator
$\tau_{SEF}^{(1)}(\varrho_{ABC}^A)$ is zero if and only if the quantum state is biseparable,
\emph{i.e.}, $\varrho_{ABC}=\sum_j p_j\rho_{AB}^j\otimes \rho_C^j+\sum_j q_j \rho_{AC}^j\otimes
\rho_B^j+\sum_j r_j \rho_A^j\otimes \rho_{BC}^j$.

When the three-qubit mixed state $\varrho_{ABC}$ is genuine tripartite entangled, its optimal pure state
decomposition contains at least one three-qubit entangled component. According to the lemma, we
obtain that $\tau_{SEF}^{(1)}(\varrho_{ABC}^A)$ is surely nonzero.

For $N$-qubit mixed states, when the indicator $\tau_{SEF}^{(1)}(\rho^{A_1}_N)$ in Eq. (12)
is zero, we can prove that there exists \emph{at most} two-qubit entanglement in the partition
$A_1|A_2\cdots A_n$ (see lemmas b and c in \cite{Suppl}) and we further have the following lemma:

\emph{Lemma 2}: In $N$-qubit mixed states, the multipartite entanglement indicator
\begin{eqnarray}\label{15}
\tau_{SEF}^{(1)}(\rho_N)=\mbox{min}\sum_jp_j \frac{\sum_{l=1}^n\tau_{SEF}^{(1)}(\ket{\psi^j}^{A_l}_N)}{N}
\end{eqnarray}
is zero if and only if the quantum state is $(N/2)$-separable in the form
$\rho_{A_1A_2\cdots A_n}=\sum_{i_1,\cdots,i_n=1}^n\sum_j p_j^{\{i_1\cdots
i_n\}}\rho_{Ai_1Ai_2}^j\otimes \cdots \otimes\rho_{Ai_{k-1}Ai_{k}}^j\otimes\cdots
\otimes\rho_{Ai_{n-1}Ai_{n}}^j$, which has at most two-qubit entanglement with the superscript
$\{i_1\cdots i_n\}$ being all permutations of the $N$ qubits.

According to lemma 2, whenever an $N$-qubit state contains genuine multiqubit entanglement,
the indicator $\tau_{SEF}^{(1)}(\rho_N)$ is surely nonzero. Thus this quantity
can serve as a genuine multiqubit entanglement indicator in
$N$-qubit mixed states. The analytical proof of this lemma and its application to an $N$-qubit
mixed state (without two-qubit concurrence and $n$-tangles) are presented in the Supplemental
Material~\cite{Suppl}.

In general, the calculation of the indicators defined in Eqs. (12) and (15) is very difficult
due to the convex roof extension. Here, based on the monogamy property of $E_f^2$ in mixed states,
we can also introduce an alternative multipartite entanglement indicator as
\begin{equation}\label{16}
\tau^{(2)}_{SEF}(\rho_N^{A_1})=E_f^2(\rho_{A_1|A_2\cdots A_n})-\sum_{j\neq1}E_f^2(\rho_{A_1A_j}),
\end{equation}
which detects the multipartite entanglement (under the given partition) not stored in pairs of qubits
(although this quantity is not monotone under local operations and classical communication
~\cite{Suppl}).
From the Koashi-Winter formula \cite{koashi04pra}, the multiqubit entanglement of formation can be
calculated by a purified state $\ket{\psi}_{A_1A_2\cdots A_nR}$ with
$\rho_{A_1A_2\cdots A_n}=\mbox{tr}_{R}\ket{\psi}\bra{\psi}$,
\begin{equation}\label{17}
E_f(A_1|A_2\cdots A_n)=D(A_1|R)+S(A_1|R),
\end{equation}
where $S(A_1|R)=S(A_1R)-S(R)$ is the quantum conditional entropy with $S(x)$ being the von Neumann
entropy, and the quantum discord $D(A_1|R)$ is defined as \cite{ollivier01prl,vedral01jpa}
\begin{equation}\label{18}
  D_{A_1|R}=\mbox{min}_{\{E_{k}^{R}\}}\sum_k p_k
  S(A_1|E_{k}^{R})-S(A_1|R)
\end{equation}
with the minimum running over all the POVMs and the measurement being performed on subsystem $R$.
Recent studies on quantum correlation provide some effective methods
\cite{luo08pra,lang10prl,gio10prl,ade10prl,ali10pra,huang13pra,cen11pra,che11pra,shi12pra}
for calculating the quantum discord, which can be used to quantify the indicator in Eq. (16).
For all partitions, we may introduce a partition-independent indicator
$\tau_{SEF}^{(2)}(\rho_N)=\sum_{i=1}^n\tau_{SEF}^{(2)}(\rho_N^{A_i})/N$.

We now apply the indicator $\tau^{(2)}_{SEF}$ to a practical dynamical procedure of a composite
system which is composed of two entangled cavity photons being affected by the dissipation of two
individual $N$-mode reservoirs. The interaction of a single cavity-reservoir
system is described by the Hamiltonian \cite{lop08prl}
$\hat{H}=\hbar \omega \hat{a}^{\dagger}\hat{a}+\hbar\sum_{k=1}^{N}\omega_{k}
\hat{b}_k^{\dagger}\hat{b}_k+\hbar\sum_{k=1}^{N}g_{k}(\hat{a}
\hat{b}_{k}^{\dagger}+\hat{b}_{k}\hat{a}^{\dagger})$.
When the initial state is $\ket{\Phi_0}=(\alpha\ket{00}+\beta\ket{11})_{c_1c_2}
\ket{00}_{r_1r_2}$ with the dissipative reservoirs being in the vacuum state,
the output state of the cavity-reservoir system has the form \cite{lop08prl}
\begin{equation}\label{19}
  \ket{\Phi_t}=\alpha\ket{0000}_{c_1r_1c_2r_2}+\beta\ket{\phi_t}_{c_1r_1}\ket{\phi_t}_{c_2r_2},
\end{equation}
where $\ket{\phi_t}=\xi(t)\ket{10}+\chi(t)\ket{01}$ with the
amplitudes being $\xi(t)=\mbox{exp}(-\kappa t/2)$ and
$\chi(t)=[1-\mbox{exp}(-\kappa t)]^{1/2}$.
As quantified by the concurrence, the entanglement
dynamical property was addressed in Refs. \cite{lop08prl,byw09pra}, but the multipartite
entanglement analysis is mainly based on some specific bipartite partitions in which each party can
be regarded as a logic qubit. When one of the parties is not equivalent to a logic qubit, the
characterization for multipartite entanglement structure is still an open problem. For example,
in the dynamical procedure, although the monogamy relation
$C^2_{c_1|c_2r_1}-C^2_{c_1c_2}-C^2_{c_1r_1}$ is satisfied, the entanglement $C^2_{c_1|c_2r_1}$ is
unavailable so far because subsystem $c_2r_1$ is a four-level system and the convex
roof extension is needed. Fortunately, in this case, we can utilize the presented indicator
$\tau^{(2)}_{SEF}(\rho_{c_1c_2r_1}^{c_1})=E_f^2(c_1|c_2r_1)-E_f^2(c_1c_2)-E_f^2(c_1r_1)$
to indicate the genuine tripartite entanglement, where $E_f(c_1|c_2r_1)$ can be obtained
via the quantum discord $D_{c_1|r_2}$~\cite{Suppl}.
This indicator detects the genuine tripartite entanglement which does not come from two-qubit
pairs. In Fig.2, the indicator and its entanglement components are plotted as functions of the time
evolution $\kappa t$ and the initial amplitude $\alpha$, where the nonzero
$\tau^{(2)}_{SEF}(c_1c_2r_1)$ actually detects the tripartite entanglement area and the
bipartite components of $E_f^2$ characterize the entanglement distribution in the dynamical
procedure. By analyzing the multipartite entanglement structure, we can know that how
the initial cavity photon entanglement transfers in the multipartite cavity-reservoir system,
which provides the necessary information to design an effective method for suppressing the decay
of cavity photon entanglement.

\begin{figure}
\begin{center}
\epsfig{figure=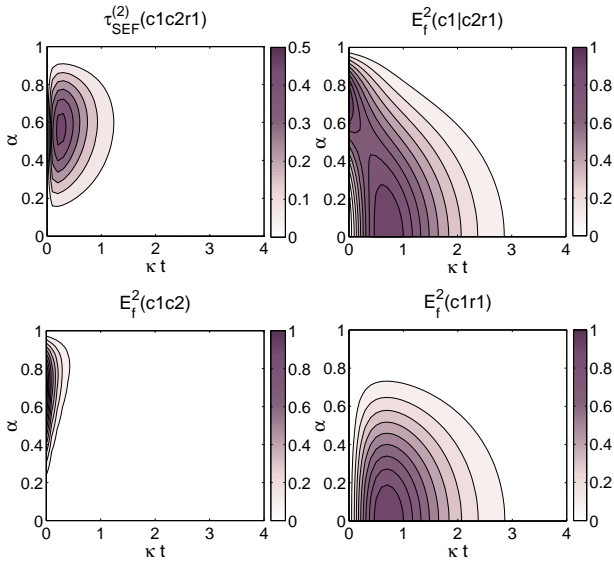,width=0.45\textwidth}
\end{center}
\caption{(Color online) The indicator $\tau_{SEF}^{(2)}(\rho_{c_1c_2r_1}^{c_1})$ and its
entanglement components as functions of the time evolution $\kappa t$ and the
initial amplitude $\alpha$, which detects the tripartite entanglement area and
illustrates the entanglement distribution in the dynamical procedure.}
\end{figure}

\emph{Discussion and conclusion.} -- The entanglement of formation is a well-defined
measure for bipartite entanglement and has the operational meaning in entanglement preparation and
data storage \cite{ben96pra}. Unfortunately,
it does not satisfy the usual monogamy relation. As an example, its monogamy score for the
three-qubit $W$ state is $E_f(A|BC)-E_f(AB)-E_f(AC)=-0.1818$. In this Letter, we show exactly that the squared
entanglement $E_f^2$ is monogamous, which mends the gap of the entanglement of formation.
Furthermore, in comparison to the monogamy of concurrence, the newly introduced indicators can
really detect all genuine multiqubit entangled states and extend the territory of entanglement
dynamics in many-body systems. In addition, via the established monogamy relation in Eq. (4),
we can obtain
\begin{equation}\label{20}
E_f(\rho_{A_1|A_2\cdots A_n})\geq \sqrt{E_f^2(\rho_{A_1A_2})+\cdots +E_f^2(\rho_{A_1A_n})},
\end{equation}
which provides a nontrivial and computable lower bound for the entanglement of formation.

In summary, we have not only proven exactly that the squared entanglement of formation
is monogamous in $N$-qubit mixed states but also provided a set of useful tool
for characterizing the entanglement in multiqubit systems, overcoming some flaws of the concurrence.
Two kinds of indicator have been introduced:
the first one can detect all genuine multiqubit entangled states and solve the critical outstanding
problem in the case of the two-qubit concurrence and $n$-tangles being zero, while the second one
can be calculated via quantum discord and applied to a practical dynamical procedure of cavity-reservoir
systems when the monogamy of concurrence loses its efficacy. Moreover, the computable lower bound
can be utilized to estimate the multiqubit entanglement of formation.

\emph{Acknowledgments.} --This work was supported by the RGC of Hong Kong under Grant Nos. HKU7058/11P 
and HKU7045/13P. Y.-K.B. and Y.-F.X. were also supported by NSF-China (Grant No. 10905016), Hebei NSF (Grant No.
A2012205062), and the fund of Hebei Normal University.

\emph{Note added.} -- Recently,  by using the same assumptions as those made in~\cite{bai13pra},
a similar idea on the monogamy of squared entanglement of formation was presented
in~\cite{oliveira13arx},  but the claimed monogamy for mixed states was not proven in that
paper~\cite{note1}, in contrast to what we have done in the present work.

\newpage

\setcounter{equation}{0}

\setcounter{figure}{0}

\section{Supplemental material}

\subsection{I. Proof of proposition I}

\emph{Proposition I}:  The squared entanglement of formation $E_f^2(C^2)$ in two-qubit mixed states
varies monotonically as a function of the squared concurrence $C^2$.

\emph{Proof}: This proposition holds if the first-order derivative $dE_f^2/dx >0$ with $x=C^2$.
According to the formula in Eq. (3) of the main text, we have
\begin{eqnarray}\label{s1}
\frac{dE_f^2}{dx}&=&-\frac{\mbox{ArcTanh}(\sqrt{1-x})}{2\sqrt{1-x}(\mbox{ln}2)^2} [2\sqrt{1-x}
\mbox{ArcTanh}(\sqrt{1-x})\nonumber\\
&&+\mbox{ln}(\frac{1-\sqrt{1-x}}{4})+\mbox{ln}(1+\sqrt{1-x})]>0.
\end{eqnarray}
The details for illustrating the positivity of Eq. (1) are as follows.

The inverse hyperbolic tangent function has the form
\begin{eqnarray}\label{s2}
\mbox{ArcTanh}(x)&=&\mbox{ln}[\sqrt{1-x^2}/(1-x)],
\end{eqnarray}
and the last two terms in Eq. (1) can be simplified as
\begin{eqnarray}\label{s3}
\mbox{ln}(\frac{1-\sqrt{1-x}}{4})+\mbox{ln}(1+\sqrt{1-x})=\mbox{ln}(\frac{x}{4}).
\end{eqnarray}
Thus the first-order derivative is
\begin{equation}\label{s4}
 \frac{dE_{f}^2}{dx}=\mathcal{T}_1\cdot\mathcal{T}_2\cdot\mathcal{T}_3,
\end{equation}
in which
\begin{eqnarray}\label{s5}
&&\mathcal{T}_1=-1/[2\sqrt{1-x}(\mbox{ln}2)^2],\nonumber\\
&&\mathcal{T}_2=\mbox{ln}[\sqrt{x}/(1-\sqrt{1-x})],\nonumber\\
&&\mathcal{T}_3=2\sqrt{1-x}\mbox{ln}[\sqrt{x}/(1-\sqrt{1-x})]+\mbox{ln}(x/4),
\end{eqnarray}
respectively. Due to $x\in(0,1)$, it is obvious that the term $\mathcal{T}_1<0$. For the term
$\mathcal{T}_2$, we have $\sqrt{x}>x$ and $\sqrt{1-x}>1-x$, which results in
\begin{eqnarray}\label{s6}
x>1-\sqrt{1-x}&\Rightarrow& \sqrt{x}>1-\sqrt{1-x}\nonumber\\
&\Rightarrow& \frac{\sqrt{x}}{1-\sqrt{1-x}}>1.
\end{eqnarray}
Therefore, the second term $\mathcal{T}_2$ in Eq. (4) is positive. For the third term, we have
\begin{eqnarray}\label{s7}
\mathcal{T}_3&=&2\sqrt{1-x}\mbox{ln}(\frac{\sqrt{x}}{1-\sqrt{1-x}})+\mbox{ln}(\frac{x}{4})\nonumber\\
&<&2\mbox{ln}(\frac{\sqrt{x}}{1-\sqrt{1-x}})+\mbox{ln}(\frac{x}{4})\nonumber\\
&=&\mbox{ln}[\frac{x/2}{1-\sqrt{1-x}}]^2<0,
\end{eqnarray}
where, in the first inequality, we use the property $\sqrt{1-x}<1$, and the last inequality is
satisfied due to $x/2<1-\sqrt{1-x}$. Since $\mathcal{T}_1<0$, $\mathcal{T}_2>0$ and
$\mathcal{T}_3<0$, the first-order derivative in Eq. (1) is positive. Combining the fact with that
$x=0$ corresponds to the minimum $E_f^2=0$ and $x=1$ corresponds to the maximum $E_f^2=1$, we get
that $E_f^2$ is a monotonically increasing function of $x$, which completes the proof of the
proposition.

\subsection{II. Proof of proposition II}

\emph{Proposition II}:  The squared entanglement of formation $E_f^2(C^2)$ is convex as a function
of the squared concurrence $C^2$.

\emph{Proof}: This proposition holds if the second-order derivative $d^2(E_f^2)/dx^2>0$. After some
deduction, we have
\begin{eqnarray}\label{s8}
\frac{d^2(E_f^2)}{dx^2}&=&g(x)\cdot \{\sqrt{1-x}\mbox{ln}(x/4)-\mbox{ArcTanh}(\sqrt{1-x})\nonumber\\
&&\times [2x-2+x\mbox{ln}(x/4)]\}>0,
\end{eqnarray}
where $g(x)=1/[4(1-x)^{3/2}x(\mbox{ln}2)^2]$ is a non-negative factor.

\begin{figure}
\begin{center}
\epsfig{figure=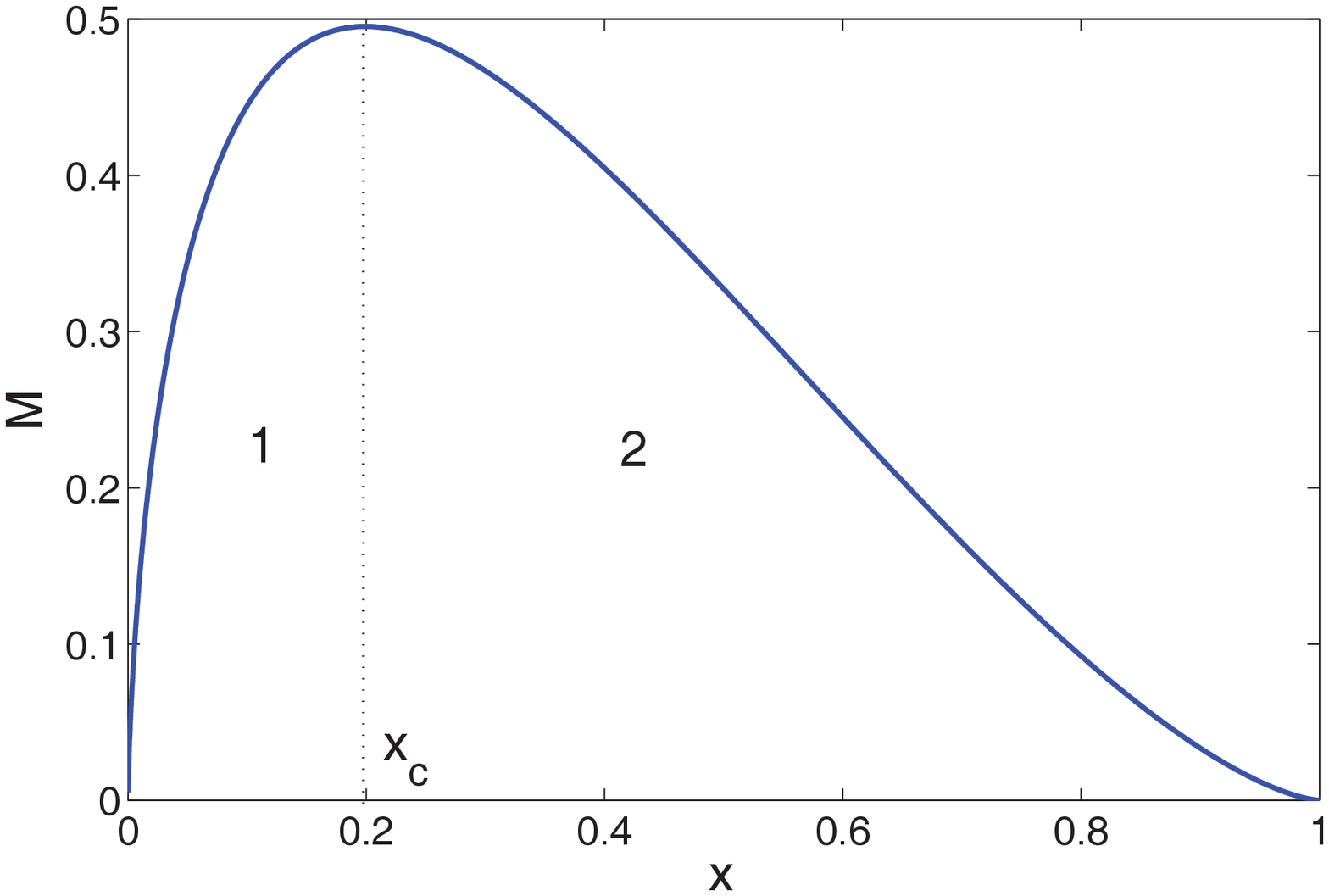,width=0.45\textwidth}
\end{center}
\caption{(Color online)The factor $\mathcal{M}$ is plotted as a function of $x$, which is
monotonically increasing in the region $x\in(0,x_c)$ and decreasing in the region $x\in(x_c,1)$
with the critical value $x_c=4/e^3\simeq 0.199$.}
\end{figure}

The detailed derivation for the above result is as follows. In Eq. (8), when the parameter $x\in
(0,1)$, the factor $g(x)=1/[4(1-x)^{3/2}x(\mbox{ln}2)^2]$ is positive. In this case, the positivity
of $d^2(E_f^2)/dx^2$ is equivalent to
\begin{equation}\label{s9}
\mathcal{M}(x)>0,
\end{equation}
where $\mathcal{M}(x)=\sqrt{1-x}\mbox{ln}(x/4)-\mbox{ArcTanh}(\sqrt{1-x})[2x-2+x\mbox{ln}(x/4)]$.
In order to analyze the sign of $\mathcal{M}$, we study its monotonic property. The first-order
derivative of $\mathcal{M}$ has the form
\begin{eqnarray}\label{s10}
\frac{d\mathcal{M}(x)}{dx}&=&-\mbox{ArcTanh}(\sqrt{1-x})[3+\mbox{ln}(\frac{x}{4})]\nonumber\\
&=&\eta_1\cdot \eta_2 ,
\end{eqnarray}
where the parameters are
\begin{eqnarray}\label{s11}
&&\eta_1=-\mbox{ln}[(1+\sqrt{1-x})/\sqrt{x}],\nonumber\\
&&\eta_2=3+\mbox{ln}(\frac{x}{4}),
\end{eqnarray}
respectively. Since $x\in (0,1)$, we have the factor $\eta_1<0$. Therefore, according to Eq. (10),
the function $\mathcal{M}(x)$ increases monotonically when the factor $\eta_2<0$ , and it decreases
monotonically for the case $\eta_2>0$. This property is shown in Fig.1, and the maximal value of
$\mathcal{M}(x)$ corresponds to the critical point $x_c=4/e^3\simeq 0.199$.

Now, we analyze two endpoints of $\mathcal{M}$ for $x=0$ and $x=1$. When $x=0$, we can deduce
\begin{eqnarray}\label{s12}
\lim_{x\to +0}\mathcal{M}(x)&=&\lim_{x\to +0}\{\sqrt{1-x}\mbox{ln}(\frac{x}{4})\nonumber\\
&&-\mbox{ln}(\frac{1+\sqrt{1-x}}{\sqrt{x}})[2x-2+x\mbox{ln}(\frac{x}{4})]\}\nonumber\\
&=&\lim_{x\to +0}[\mbox{ln}(\frac{x}{4})-\mbox{ln}(\frac{1+\sqrt{1-x}}{\sqrt{x}})\times (-2)]\nonumber\\
&=&\lim_{x\to +0}\mbox{ln}[\frac{(1+\sqrt{1-x})^2}{4}]\nonumber\\
&=&0,
\end{eqnarray}
where in the second equation we have used the property $\lim_{x\to +0}x\mbox{ln}(x/4)=0$. For the
other endpoint $x=1$, we can get $\mathcal{M}(1)=0$. Combining it with the monotonic properties of
$\mathcal{M}$, we can find $\mathcal{M}(x)>0$, which is equivalent to $d^2(E_f^2)/dx^2>0$ in the
region $x\in(0,1)$.

Furthermore, we analyze the value of second-order derivative $d^2(E_f^2)/dx^2$ at the endpoints.
When $x=0$, we can get
\begin{eqnarray}\label{s13}
\lim_{x\to +0}\frac{d^2(E_{f}^2)}{dx^2}&=&\lim_{x\to +0} g(x)\cdot \mathcal{M}(x)\nonumber\\
&=&\lim_{x\to +0}\frac{-\mbox{ln}(\frac{1+\sqrt{1-x}}{\sqrt{x}})
[3+\mbox{ln}(\frac{x}{4})]}{2(2-5x)\sqrt{1-x}(\mbox{ln}2)^2}\nonumber\\
&=&\frac{1}{4(\mbox{ln}2)^2}\lim_{x\to +0}\{[\mbox{ln}(\frac{4}{x})-3]\nonumber\\
&&\times\mbox{ln}(\frac{1+\sqrt{1-x}}{\sqrt{x}})\}\nonumber\\
&=&\infty.
\end{eqnarray}
On the other hand, when $x=1$, we have
\begin{eqnarray}\label{s14}
&&\lim_{x\to 1}\frac{d^2(E_{f}^2)}{dx^2}\nonumber\\
&=&\lim_{x\to 1}\frac{[\mbox{ln}(\frac{4}{x})-3]\mbox{ln}(\frac{1+\sqrt{1-x}}{\sqrt{x}})}
{2(2-5x)\sqrt{1-x}(\mbox{ln}2)^2}\nonumber\\
&=&\lim_{x\to1}[\frac{3-\mbox{ln}4-\mbox{ln}(\frac{1}{x})-2\sqrt{1-x}\mbox{ln}
(\frac{1+\sqrt{1-x}}{\sqrt{x}})}{6x(5x-4)(\mbox{ln}2)^2}]\nonumber\\
&=&\frac{3-\mbox{ln}4}{6\times(\mbox{ln}2)^2}\nonumber\\
&\approx& 0.55979.
\end{eqnarray}
Thus, we have shown the second-order derivative $d^2(E_f^2)/dx^2>0$ in the whole region
$x\in[0,1]$, and then complete the proof of proposition II. In Fig.2, the derivative is plotted as
a function of $x$, which illustrates our result.

\begin{figure}
\begin{center}
\epsfig{figure=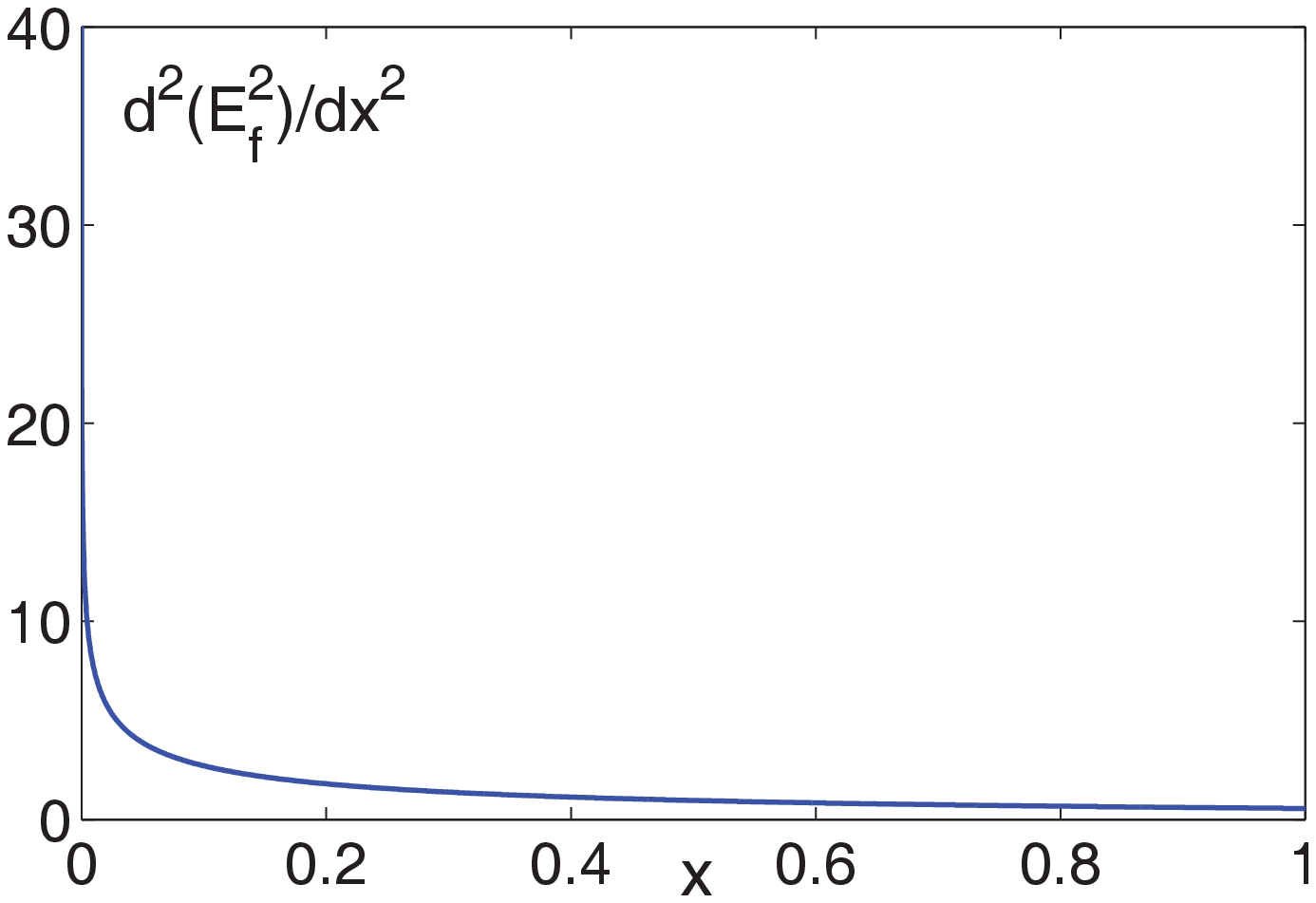,width=0.45\textwidth}
\end{center}
\caption{(Color online) The second-order derivative $d^2(E_f^2)/dx^2$ is plotted as a function of
$x$, which is positive and the two endpoints value are $\infty$ and $0.55979$, respectively.}
\end{figure}

\subsection{III. Proof for the inequalities in Eq. (5)}

According to the proposition $I$ in the Letter, we know that the squared entanglement of formation
$E_f^2(C^2)$ is monotonically increasing as a function of the squared concurrence $C^2$. Combining
this property with the monogamy relation of concurrence in Eq. (1) of the main text, we can derive
the first inequality in Eq. (5) of Letter
\begin{eqnarray}\label{s15}
E_f^2(C^2_{A_1|A_2\cdots A_n})\geq E_f^2(C_{A_1A_2}^2+\cdots +C_{A_1A_n}^2).
\end{eqnarray}
Here, it should be emphasized that the composite system $A_1A_2\cdots A_n$ is in a pure state
$\ket{\psi}_{A_1A_2\cdots A_n}$. The reason is that, in a generic mixed state $\rho_{A_1A_2\cdots
A_n}$, the relation between the entanglement of formation $E_f(A_1|A_2\cdots A_n)$ and the squared
concurrence $C^2(A_1|A_2\cdots A_n)$ can not be characterized by Eq. (3) of Letter since the
subsystem $A_2\cdots A_n$ is not equivalent to a logic qubit.

\begin{figure}
\begin{center}
\epsfig{figure=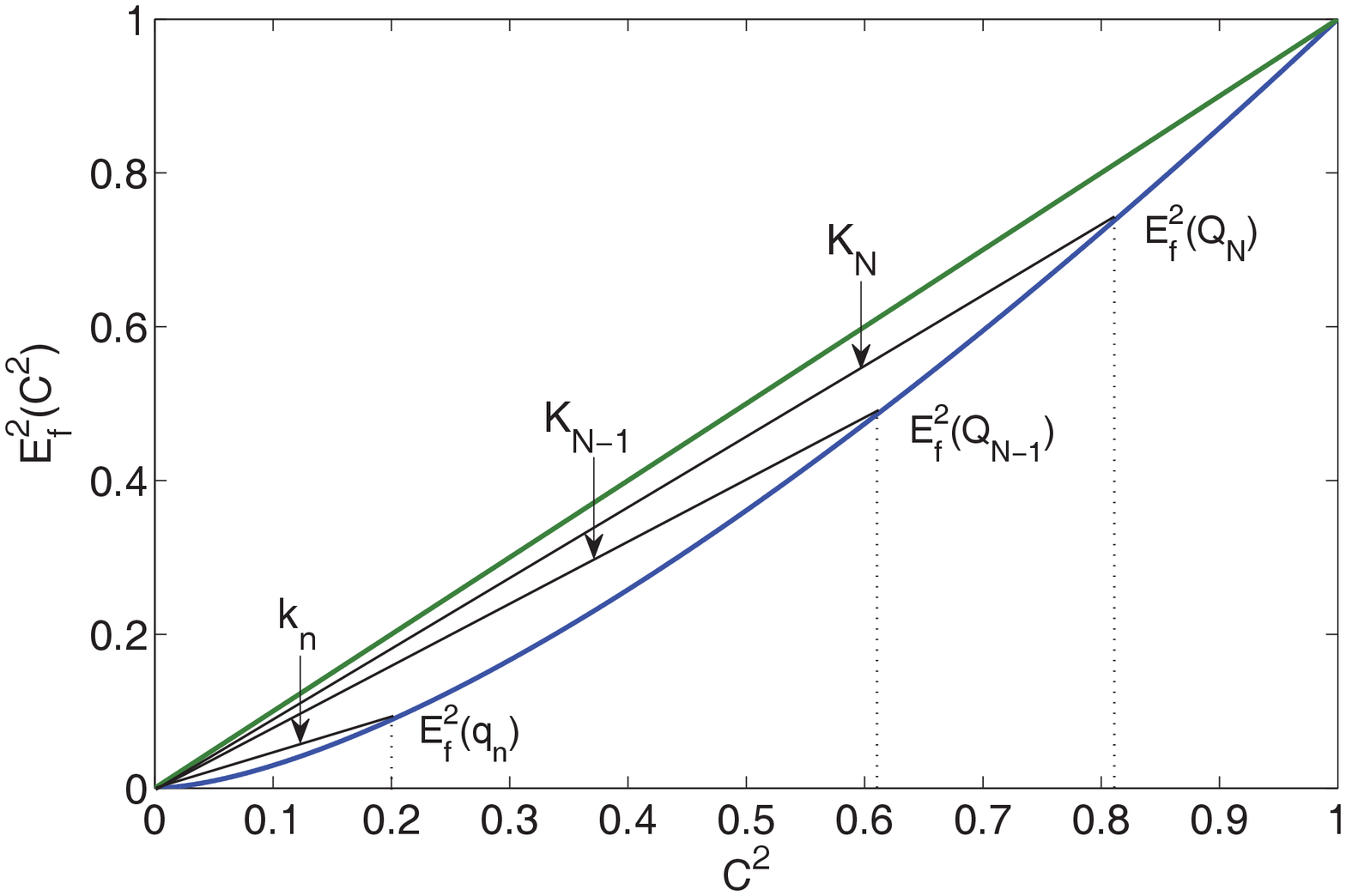,width=0.45\textwidth}
\end{center}
\caption{(Color online) A schematic diagram for the monogamy inequality in Eq. (16). The blue line
is the squared entanglement of formation $E_f^2(C^2)$ and the green line is the squared concurrence
$C^2$, where the gradients $K_N>K_{N-1}$ and $K_N>k_n$ with $Q_N=\sum_{i=2}^{n}C_{A_1A_i}^2$,
$Q_{N-1}=\sum_{i=2}^{n-1}C_{A_1A_i}^2$, and $q_n=C_{A_1A_n}^2$, respectively.}
\end{figure}

Furthermore, according to the proposition $II$ that the squared entanglement of formation
$E_f^2(C^2)$ is convex as a function of $C^2$, we can derive
\begin{eqnarray}\label{s16}
&& E_f^2(C_{A_1A_2}^2+\cdots +C_{A_1A_n}^2)\nonumber\\
&\geq& E_f^2(C_{A_1A_2}^2+\cdots +C_{A_1A_{n-1}}^2)+E_f^2(C_{A_1A_n}^2).
\end{eqnarray}
A schematic diagram for this inequality is shown in Fig.3. Due to the monotonic and convex property
of $E_f^2(C^2$), we have the relations of gradients $K_N>K_{N-1}$ and $K_N>k_n$, which give rise to
\begin{eqnarray}\label{s17}
&&E^2_f(Q_N)-E^2_f(Q_{N-1})-E_f^2(q_n)\nonumber\\
&=&K_{N}Q_N-K_{N-1}Q_{N-1}-k_nq_n\nonumber\\
&\geq& K_{N}(Q_N-Q_{N-1}-q_n)\nonumber\\
&=&0,
\end{eqnarray}
where $Q_N=\sum_{i=2}^{n}C_{A_1A_i}^2$, $Q_{N-1}=\sum_{i=2}^{n-1}C_{A_1A_i}^2$, and
$q_n=C_{A_1A_n}^2$, respectively. By iterating the property of gradients $K_{i}$ and $k_{j}$, we
finally have the inequality
\begin{eqnarray}\label{s18}
&& E_f^2(C_{A_1A_2}^2+\cdots +C_{A_1A_n}^2)\nonumber\\
&\geq& E_f^2(C_{A_1A_2}^2)+E_f^2(C_{A_1A_3}^2)+\cdots+E_f^2(C_{A_1A_n}^2).
\end{eqnarray}
Combining Eqs. (18) and  (15), we can obtain the inequalities in Eq. (5) of Letter.

\subsection{IV. The second term in Eq. (9) being non-negative}

In the Letter, the second term in Eq. (9) has the form
\begin{equation}\label{s19}
2\sum_{i}\sum_{k=i+1}(E1_{i}E1_{k}-\sum_{j}Ej_{i}Ej_{k}),
\end{equation}
and here we prove that it is non-negative. For any pure state component $\ket{\psi^i}$ in Eq. (7)
of the main text, the monogamy property of $E_f^2$ is satisfied, and we have
\begin{equation}\label{s20}
E1_{i}^2\geq \sum_{j=2}^{n}Ej_{i}^2.
\end{equation}
For the two arbitrary pure state components $\ket{\psi^i}$ and $\ket{\psi^k}$, we can get
\begin{eqnarray}\label{s21}
E1_{i}^2E1_{k}^2&\geq& (\sum_{j=2}^{n}Ej_{i}^2)(\sum_{p=2}^{n}Ep_{k}^2)\nonumber\\
&=&\sum_{j=2}^{n}(Ej_{i}Ej_{k})^2\nonumber\\
&&+\sum_{p=2}^{n-1}\sum_{q=p+1}^{n}(Ep_{i}^2Eq_{k}^2+Eq_{i}^2Ep_{k}^2)\nonumber\\
&\geq& \sum_{j=2}^{n}(Ej_{i}Ej_{k})^2\nonumber\\
&&+\sum_{p=2}^{n-1}\sum_{q=p+1}^{n}(2Ep_{i}Ep_{k}Eq_{i}Eq_{k})\nonumber\\
&=&(\sum_{j}Ej_{i}Ej_{k})^2 ,
\end{eqnarray}
where in the second inequality we have used the relation of the perfect square trinomials
\begin{equation}\label{s22}
Ep_{i}^2Eq_{k}^2+Eq_{i}^2Ep_{k}^2\geq 2Ep_{i}Ep_{k}Eq_{i}Eq_{k}.
\end{equation}
After taking the square root on both sides of Eq. (21), we have
\begin{eqnarray}\label{s23}
&& E1_{i}E1_{k}\geq \sum_{j}Ej_{i}Ej_{k}\nonumber\\
&& \Rightarrow E1_{i}E1_{k}-\sum_{j}Ej_{i}Ej_{k}\geq 0.
\end{eqnarray}
Since $\ket{\psi^i}$ and $\ket{\psi^k}$ are two arbitrary components in Eq. (7) of the Letter, the
above relation of Eq. (23) is also satisfied for any other components. Therefore, we can obtain
\begin{equation}\label{s24}
2\sum_{i}\sum_{k=i+1}(E1_{i}E1_{k}-\sum_{j}Ej_{i}Ej_{k})\geq 0,
\end{equation}
which completes our proof.

\subsection{V. Proof of the lemma 1}

In order to  prove the lemma 1 presented in the main text, we here first prove the following lemma.

\emph{Lemma a}. For three-qubit pure states, the multipartite entanglement indicator
$\tau_{SEF}^{(1)}(\ket{\psi_{A|BC}})$ is zero if and only if the quantum state is a bipartite
product state, i.e., $\ket{\psi_{ABC}}=\ket{\phi_i}\otimes\ket{\phi_j}$ with $i, j\in\{A, B, C\}$.

\emph{Proof}. We first prove the necessity. When the quantum state is bipartite product, the
three-qubit state has the forms:
\begin{eqnarray}\label{s25}
&&\ket{\psi_{ABC}}=\ket{\phi_{AB}}\otimes \ket{\phi_C}\nonumber\\
&&\ket{\psi_{ABC}}=\ket{\phi_{AC}}\otimes \ket{\phi_B}\nonumber\\
&&\ket{\psi_{ABC}}=\ket{\phi_{A}}\otimes \ket{\phi_{BC}}\nonumber\\
&&\ket{\psi_{ABC}}=\ket{\phi_{A}}\otimes \ket{\phi_B}\otimes \ket{\phi_C}.
\end{eqnarray}
It is easy to obtain $\tau_{SEF}^{(1)}(\ket{\psi_{A|BC}})=E_f^2(\ket{\psi_{A|BC}})-E_f^2(\rho_{AB})
-E_f^2(\rho_{AC})=0$ for these product states.

We next show the sufficiency. It is a fact that when the indicator
$\tau_{SEF}^{(1)}(\ket{\psi_{A|BC}})$ is zero, there is \emph{at most} one nonzero two-qubit
concurrence in the three-qubit pure state. This is because that, if $C^2_{AB}>0$ and $C^2_{AC}>0$,
we have
\begin{eqnarray}\label{s26}
&&E_f^2(C^2_{A|BC})-E_f^2(C^2_{AB})-E_f^2(C^2_{AC})\nonumber\\
&\geq& E_f^2(C^2_{AB}+C^2_{AC})-E_f^2(C^2_{AB})-E_f^2(C^2_{AC})\nonumber\\
&=& k_{A|BC}(C^2_{AB}+C^2_{AC})-k_{AB}C^2_{AB}-k_{AC}C^2_{AC}\nonumber\\
&>&0,
\end{eqnarray}
which is contradictory to the premise $\tau_{SEF}^{(1)}=0$. Here, in the first inequality we use
proposition $I$, and in the third inequality we use proposition $II$ with $k_i=E_f^2/C^2$ (as shown
in Fig.3) which results in the nonzero value of the indicator in the case of $C^2_{AB}>0$ and
$C^2_{AC}>0$. When both the two-qubit concurrences are zero, we have the entanglement
$E_f^2(A|BC)=0$, which corresponds to the product states $\ket{\psi_{ABC}}=\ket{\phi_{A}}\otimes
\ket{\phi_{BC}}$ and $\ket{\psi_{ABC}}=\ket{\phi_{A}}\otimes \ket{\phi_B}\otimes\ket{\phi_C}$.

In the following, we will prove that the three-qubit pure state is a bipartite product state when
the indicator is zero with one nonzero two-qubit concurrence. Without loss of generality, we assume
that the concurrence $C_{AB}$ is nonzero. In this case, we can obtain
\begin{equation}\label{s27}
E_f(\ket{\psi_{A|BC}})=E_f(\rho_{AB}).
\end{equation}
According to the definition of the entanglement of formation \cite{horo09rmp-s,wootters98prl-s}, we
further have
\begin{equation}\label{s28}
S(\rho_A)=\mbox{min}\sum_i p_i S(\rho_A^i),
\end{equation}
where $\rho_A$ is the reduced density matrix of subsystem $A$ and the minimum runs over all the
pure state decompositions $\rho_{AB}=\sum_i p_i \proj{\phi_{AB}^i}$ with
$\rho_A^i=\mbox{tr}_B(\proj{\phi_{AB}^i})$. On the other hand, we known that the von Neumann
entropy is a strictly concave function of its input \cite{nie00book-s},
\begin{equation}\label{s29}
S(\rho_A)\geq \sum_i p_i S(\rho_A^i),
\end{equation}
where the reduced density matrix $\rho_A=\sum_i p_i\rho_A^i$. Note that the equality holds if and
only if all the state $\rho_A^i$ are identical \cite{nie00book-s}. Combining Eqs. (28) and (29), we
can obtain that, if $\rho_{AB}$ is a mixed state, all the pure state components $\ket{\phi_{AB}^i}$
in an arbitrary pure state decomposition should have the identical reduced state
\begin{equation}\label{s30}
\rho_A^i=\rho_A.
\end{equation}
Now we prove that $\rho_{AB}$ cannot be a mixed state by analyzing the generic form of three-qubit
pure states. Under local unitary operations, the standard form of three-qubit pure state can be
written as \cite{acin00prl-s}
\begin{eqnarray}\label{s31}
  \ket{\psi_{ABC}}&=&\lambda_0
  \ket{000}+\lambda_1e^{i\varphi}\ket{100}+\lambda_2\ket{101}+\lambda_3\ket{110}\nonumber\\
  &&+\lambda_4\ket{111},
\end{eqnarray}
where the real number $\lambda_i$ ranges in $[0,1]$ with the condition $\sum \lambda_i^2=1$, and
the relative phase $\varphi$ changes in $[0,\pi]$. Its two-qubit reduced state of subsystem $AB$
can be expressed as
\begin{equation}\label{s32}
\varrho_{AB}=\proj{\phi_{AB}^1}+\proj{\phi_{AB}^2}
\end{equation}
with the non-normalized pure state components being
\begin{eqnarray}\label{s33}
&&\ket{\phi_{AB}^{1}}=\lambda_0 \ket{00}+\lambda_1 e^{i\varphi}\ket{10}+\lambda_3\ket{11},\nonumber\\
&&\ket{\phi_{AB}^{2}}=\lambda_2\ket{10}+\lambda_4\ket{11}.
\end{eqnarray}
From the requirement in Eq. (30) plus the nonzero value of $S(\rho_A)$ in the premise, we conclude
that for both components the entropy should satisfy $S(\rho_A^{i})=S(\rho_A)>0$ . However, in Eq.
(33), the second component $\ket{\phi_{AB}^{2}}$ is a product state and we have $S(\rho_A^{(2)})=0$
due to $\rho_A^{(2)}=\proj{1}$. This means that the probability for $\ket{\phi_{AB}^{2}}$ is zero
and then $\rho_{AB}$ can be written as
\begin{equation}\label{s34}
\rho_{AB}=\proj{\phi_{AB}^{1}}
\end{equation}
being a two-qubit pure state. Therefore, in three-qubit pure states, when the indicator
$\tau_{SEF}^{(1)}=0$ with the nonzero concurrence $C_{AB}$ the composite system has the form
\begin{equation}\label{s35}
\ket{\psi_{ABC}}=\ket{\phi_{AB}}\otimes \ket{\phi_C}.
\end{equation}
Similarly, for the case $C_{AC}>0$, we can derive that $\ket{\psi_{ABC}}=\ket{\phi_{AC}}\otimes
\ket{\phi_B}$. This completes the proof of the lemma in three-qubit pure states.

At this stage, we prove the lemma 1 of the Letter, which is stated as:

\emph{Lemma 1}. For three-qubit mixed states, the multipartite entanglement indicator
$\tau_{SEF}^{(1)}(\varrho_{ABC}^A)$ is zero if and only if the quantum state is biseparable,
\emph{i.e.}, $\varrho_{ABC}=\sum_j p_j\rho_{AB}^j\otimes \rho_C^j+\sum_j q_j \rho_{AC}^j\otimes
\rho_B^j+\sum_j r_j \rho_A^j\otimes \rho_{BC}^j$.

\emph{Proof}. For three-qubit mixed states, the tripartite entanglement indicator is defined as
\begin{equation}\label{s36}
\tau_{SEF}^{(1)}(\varrho_{ABC}^A)=\mbox{min}\sum_i p_i \tau_{SEF}^{(1)}(\ket{\psi_{A|BC}^i}),
\end{equation}
where the minimum runs over all the pure state decompositions $\{p_i,\ket{\psi^i_{ABC}}\}$. For the
biseparable state in the lemma, we can obtain $\tau_{SEF}^{(1)}=0$, which proves the necessity of
the lemma.  Next, we analyze the sufficiency of the lemma. When the indicator is zero, according to
the previous lemma a in three-qubit pure states, we known that there must exist an optimal pure
state decomposition for $\varrho_{ABC}$ in which every pure state component is bipartite product
and has the forms shown in Eq. (25). In the general case, this kind of three-qubit mixed state can
be written as
\begin{eqnarray}\label{s37}
\varrho_{ABC}&=&\sum_j p_j\rho_{AB}^j\otimes \rho_C^j+\sum_j q_j \rho_{AC}^j\otimes \rho_B^j\nonumber\\
&&+\sum_j r_j \rho_A^j\otimes \rho_{BC}^j
\end{eqnarray}
which is just the biseparable entangled state and does not contain the genuine tripartite
entanglement \cite{huber10prl-s}. This completes the proof for the lemma.

\subsection{VI. Proof of lemma 2 and its application to an $N$-qubit mixed
state without two-qubit concurrence and $n$-tangles}

In order to prove the lemma $2$ presented in the main text, we first prove the following two
lemmas.

\emph{Lemma b}. In $N$-qubit pure states, the multipartite entanglement indicator
$\tau_{SEF}^{(1)}(\ket{\psi}^{A_1}_N)$ is zero if and only if the quantum state is a bipartite
product state in the forms
\begin{eqnarray}\label{s38}
&&\ket{\psi}_{N}=\ket{\phi}_{A_1}\otimes\ket{\phi}_{(\overline{A_1})_{N-1}}\nonumber\\
&&\ket{\psi}_{N}=\ket{\phi}_{A_1A_k}\otimes\ket{\phi}_{(\overline{A_1A_k})_{N-2}},
\end{eqnarray}
where $\overline{A_1}$ and $\overline{A_1A_k}$ are the complementary sets of qubits $A_1$ and
$A_1A_k$ with $k\in \{2,3,\cdots ,n\}$, respectively.

\emph{Proof}. We first prove the necessity. When the quantum state is bipartite separable in the
forms shown in Eq. (38), it is easy to obtain
$\tau_{SEF}^{(1)}(\ket{\psi}_N^{A_1})=E_f^2(\ket{\psi}_{A_1|(\overline{A_1})_{N-1}}) -\sum_{k=2}^n
E_f^2(\rho_{A_1A_k})=0$ for the two kinds of product states.

We next show the sufficiency. When the entanglement indicator
$\tau_{SEF}^{(1)}(\ket{\psi}_N^{A_1})$ is zero, we can obtain that there is \emph{at most} one
nonzero two-qubit concurrence $C_{A_1A_k}$, which is due to the monotonic and strictly convex
property of $E_f^2$ as a function of squared concurrence $C^2$ (propositions $I$ and $II$). If two
concurrences are nonzero, the indicator is inevitably nonzero (similar to the three-qubit case in
Eq. 26), which is contradictory to the premise of zero indicator value. In the case of all
two-qubit concurrences $C_{A_1A_k}^2$ ($k=2,\cdots, n$) being zero, we have the entanglement
$E_f^2(A_1|A_2\cdots A_n)=0$ and then the $N$-qubit pure state has the form
$\ket{\psi_{N}}=\ket{\phi}_{A_1}\otimes\ket{\phi}_{(\overline{A_1})_{N-1}}$.

In the following, we will prove that the $N$-qubit pure state has the form
$\ket{\psi_{N}}=\ket{\phi}_{A_1A_k}\otimes\ket{\phi}_{(\overline{A_1A_k})_{N-2}}$ when the
indicator is zero with one nonzero concurrence $C_{A_1A_k}$. Without loss of generality, we assume
that the concurrence $C_{A_1A_2}$ is nonzero. In this case, we have $E_f(\ket{\psi}_{A_1|A_2\cdots
A_n})=E_f(\rho_{A_1A_2})$ and $S(\rho_{A_1})=\mbox{min}\sum_i p_i S(\rho_{A_1}^i)$. Due to the
strictly concave property of the von Neumann entropy \cite{nie00book-s}, we obtain that when
$\rho_{A_1A_2}$ is a mixed state, all its pure state components $\ket{\phi^i}_{A_1A_2}$ in an
arbitrary pure state decomposition should have the identical reduced state
\begin{equation}\label{s39}
\rho_{A_1}^i=\rho_{A_1}.
\end{equation}

Next, we prove that $\rho_{A_1A_2}$ cannot be a mixed state. We assume that, in the spectral
decomposition, the two-qubit mixed state can be written as
\begin{eqnarray}\label{s40}
\rho_{A_1A_2}=\sum_{i=1}^4 p_i\proj{\varphi^i_{A_1A_2}},
\end{eqnarray}
where the reduced state of each orthogonal component has the same form
$\rho_{A_1}^i=\mbox{tr}_{A_2}[\proj{\varphi^i_{A_1A_2}}]=\rho_{A_1}$ with $i=1,2,3,4$,
respectively. According to the reduction interpretation of $\rho_{A_1A_2}$, the mixed state comes
from a larger pure state via the partial trace of environment subsystem. Different measurement on
the environment subsystem results in different pure state decomposition. Thus, for the $N$-qubit
pure state $\ket{\psi}_N$, the qubits $A_3A_4\cdots A_n$ are the environment subsystem and
equivalent to two logic qubits $E_1E_2$. The spectral decomposition of $\rho_{A_1A_2}$ in Eq. (40)
can be obtained by the measurement $\{\ket{00},\ket{01},\ket{10},\ket{11}\}$ on logic qubits
$E_1E_2$, and the enlarged pure state can be expressed as
\begin{eqnarray}\label{s41}
\ket{\psi}_{A_1A_2E_1E_2}&=&\sqrt{p_1}\ket{\varphi^1_{A_1A_2}}\ket{00}_{E_1E_2}\nonumber\\
&+&\sqrt{p_2}\ket{\varphi^2_{A_1A_2}}\ket{01}_{E_1E_2}\nonumber\\
&+&\sqrt{p_3}\ket{\varphi^3_{A_1A_2}}\ket{10}_{E_1E_2}\nonumber\\
&+&\sqrt{p_4}\ket{\varphi^4_{A_1A_2}}\ket{11}_{E_1E_2}.
\end{eqnarray}
When we measure the logic qubit $E_2$ on the basis $\{\ket{0}_{E_2},\ket{1}_{E_2}\}$, the mixed
state of subsystem $A_1A_2E_1$ can be written as
\begin{eqnarray}\label{s42}
\rho_{A_1A_2E_1}=q_0\proj{\Phi^0}+q_1\proj{\Phi^1},
\end{eqnarray}
where the probabilities are $q_0=p_1+p_3$ and $q_1=p_2+p_4$, and two components have the forms
\begin{eqnarray}
&&\ket{\Phi^0}=a_0\ket{\varphi^1_{A_1A_2}}\ket{0}_{E_1}+b_0\ket{\varphi^3_{A_1A_2}}\ket{1}_{E_1}\\
&&\ket{\Phi^1}=a_1\ket{\varphi^2_{A_1A_2}}\ket{0}_{E_1}+b_1\ket{\varphi^4_{A_1A_2}}\ket{1}_{E_1}
\end{eqnarray}
with $a_0=\sqrt{p_1/q_0}$, $b_0=\sqrt{p_3/q_0}$, $a_1=\sqrt{p_2/q_1}$, and $b_1=\sqrt{p_4/q_1}$,
respectively. The first component $\ket{\Phi^0}$ is equivalent to a three-qubit pure state, and,
under local unitary transformation, it can be expressed as \cite{acin00prl-s}
\begin{eqnarray}\label{45}
\ket{\Phi^0}_{A_1A_2E_1}&=&\lambda_0
  \ket{\bar{0}\bar{0}\bar{0}}+\lambda_1e^{i\theta}\ket{\bar{1}\bar{0}\bar{0}}
  +\lambda_2\ket{\bar{1}\bar{0}\bar{1}}\nonumber\\
  &&+\lambda_3\ket{\bar{1}\bar{1}\bar{0}}+\lambda_4\ket{\bar{1}\bar{1}\bar{1}}.
\end{eqnarray}
When we measure the logic qubit $E_1$ in the basis $\{\ket{\bar{0}}_{E_1},\ket{\bar{1}}_{E_1}\}$,
the two pure state components of $A_1A_2$ are
\begin{eqnarray}\label{s46}
&&\ket{\bar{\varphi}_{A_1A_2}^1}=\lambda_0 \ket{\bar{0}\bar{0}}+\lambda_1
e^{i\theta}\ket{\bar{1}\bar{0}}+\lambda_3\ket{\bar{1}\bar{1}},\nonumber\\
&&\ket{\bar{\varphi}_{A_1A_2}^3}=\lambda_2\ket{\bar{1}\bar{0}}+\lambda_4\ket{\bar{1}\bar{1}}.
\end{eqnarray}
Due to the component $\ket{\bar{\varphi}_{A_1A_2}^3}$ being a product state, its probability is
zero according to the condition in Eq. (39) which requires that each component in an arbitrary pure
state decomposition should have the equal entanglement. Therefore, the three-qubit state
$\ket{\Phi^0}_{A_1A_2E_1}$ can be rewritten in the form
\begin{eqnarray}\label{s47}
\ket{\Phi^0}_{A_1A_2E_1}=\ket{\bar{\varphi}_{A_1A_2}^1}\otimes \ket{\bar{0}}_{E_1}.
\end{eqnarray}
Combining Eqs. (47) and (43), we can obtain that, for the two components $\ket{\varphi^1_{A_1A_2}}$
and $\ket{\varphi^3_{A_1A_2}}$ in the spectral decomposition of $\rho_{A_1A_2}$, there exists only
\emph{one}. This is because the three-qubit state $\ket{\Phi^0}_{A_1A_2E_1}$ is a product state in
the partition $A_1A_2|E_1$ according to Eq. (47). If both the two components
$\ket{\varphi^1_{A_1A_2}}$ and $\ket{\varphi^3_{A_1A_2}}$ exist, then the state
$\ket{\Phi^0}_{A_1A_2E_1}$ in Eq. (43) will be an entangled state which is contradictory to the
expression in Eq. (47). Without loss of generality, we assume that the component
$\ket{\varphi^1_{A_1A_2}}$ exists. Similarly, for the three-qubit state $\ket{\Phi^1}$ in Eq. (44),
we can obtain that there is only one component between $\ket{\varphi^2_{A_1A_2}}$ and
$\ket{\varphi^4_{A_1A_2}}$. Here, we assume the component $\ket{\varphi^2_{A_1A_2}}$ exists. In
this case, the two-qubit state $\rho_{A_1A_2}$ in Eq. (40) is a rank-2 state, and has \emph{at
most} two components $\ket{\varphi^1_{A_1A_2}}$ and $\ket{\varphi^2_{A_1A_2}}$ in its spectral
decomposition. Thus, the four-qubit state in Eq. (41) is equivalent to a three-qubit state and can
be rewritten as
\begin{eqnarray}\label{s48}
\ket{\psi}_{A_1A_2E}=\sqrt{p_1}\ket{\varphi^1_{A_1A_2}}\ket{0}_{E}+
\sqrt{p_2}\ket{\varphi^2_{A_1A_2}}\ket{1}_{E},
\end{eqnarray}
where the logic qubit $E$ represents the environment subsystem $A_3A_4\cdots A_n$. When the
entanglement indicator $\tau_{SEF}^{(1)}(\ket{\psi}_N^{A_1})$ is zero with the nonzero
$C_{A_1A_2}$, the quantum state $\ket{\psi}_{A_1A_2E}$ has the following form according to the
previous lemma a
\begin{eqnarray}\label{s49}
\ket{\psi}_{A_1A_2E}=\ket{\varphi_{A_1A_2}}\otimes\ket{\varphi}_{E}.
\end{eqnarray}
Note that the forms
\begin{eqnarray}\label{s50}
&&\ket{\psi}_{A_1A_2E}=\ket{\varphi_{A_1}}\otimes\ket{\varphi_{A_2}}\otimes\ket{\varphi}_{E}\nonumber\\
&&\ket{\psi}_{A_1A_2E}=\ket{\varphi_{A_1}}\otimes\ket{\varphi_{A_2E}}\nonumber\\
&&\ket{\psi}_{A_1A_2E}=\ket{\varphi_{A_1E}}\otimes\ket{\varphi_{A_2}}
\end{eqnarray}
are excluded since they are contradictory to the condition that $C_{A_1A_2}$ is nonzero. Combining
Eqs. (49) and (48), we can obtain that the two-qubit quantum state $\rho_{A_1A_2}$ has only one
component in its spectral decomposition, and then it is a two-qubit pure state
$\ket{\phi}_{A_1A_2}$. Therefore, when the indicator $\tau_{SEF}^{(1)}(\ket{\psi}_N^{A_1})$ is zero
with a nonzero $C_{A_1A_2}$, the $N$-qubit pure state has the form
\begin{equation}\label{s51}
\ket{\psi}_{N}=\ket{\phi}_{A_1A_2}\otimes\ket{\phi}_{(\overline{A_1A_2})_{N-2}}.
\end{equation}
Similarly, when the nonzero concurrence is $C_{A_1A_k}$ with $k\in\{2,3,4,\cdots, n\}$, we can get
the $N$-qubit pure state
\begin{equation}\label{s52}
\ket{\psi}_{N}=\ket{\phi}_{A_1A_k}\otimes\ket{\phi}_{(\overline{A_1A_k})_{N-2}},
\end{equation}
which completes the proof of lemma b.

\emph{Lemma c}. In $N$-qubit mixed states, the multipartite entanglement indicator
$\tau_{SEF}^{(1)}(\rho_{N}^{A_1})$ is zero if and only if the quantum state is biseparable in the
form
\begin{eqnarray}\label{s53}
\rho_{A_1A_2\cdots A_n}&=&\sum_j p_j^1\cdot\rho_{A_1}^j\otimes
\rho_{(\overline{A_1})_{N-1}}^j\nonumber\\
&&+\sum_{k=2}^n\sum_j p_j^k\cdot\rho_{A_1A_k}^j\otimes \rho_{(\overline{A_1A_k})_{N-2}}^j,
\end{eqnarray}
where $\overline{A_1}$ and $\overline{A_1A_k}$ are the complementary sets of qubits $A_1$ and
$A_1A_k$, respectively.

\emph{Proof}. In $N$-qubit mixed states, the multipartite entanglement indicator is defined as
\begin{eqnarray}\label{s54}
\tau_{SEF}^{(1)}(\rho_{N}^{A_1})&=&\mbox{min}\sum_i
p_i[E_f^2(\ket{\psi^i}_{A_1|(\overline{A_1})_{N-1}})\nonumber\\
&&-\sum_{k=2}^n E_f^2(\rho^i_{A_1A_k})],
\end{eqnarray}
where the minimum runs over all the pure state decomposition $\{p_i, \ket{\psi_N^i}\}$. For the
biseparable state in Eq. (53), it is easy to obtain that the indicator
$\tau_{SEF}^{(1)}(\rho_{N}^{A_1})$ is zero, which proves the necessity of the lemma. Next, we
consider the sufficiency of the lemma. When the entanglement indicator
$\tau_{SEF}^{(1)}(\rho_{N}^{A_1})$ being zero, according to the previous lemma b in $N$-qubit pure
state, we can obtain that there must exist an optimal pure state decomposition for $\rho_{N}$ in
which each pure state component is bipartite product and has the forms shown in Eq. (38). In the
general case, this kind of $N$-qubit mixed state can be written as the form shown in Eq. (53),
which completes the proof of lemma c.

According to lemmas b and c, we can obtain that, when the indicator
$\tau_{SEF}^{(1)}(\rho^{A_1}_N)$ is zero, there is \emph{at most} two-qubit entanglement in the
partition $A_1|A_2\cdots A_n$. On the other hand, whenever the multiqubit entanglement exists in
this partition $A_1|A_2\cdots A_n$, the entanglement indicator $\tau_{SEF}^{(1)}(\rho^{A_1}_N)$ is
surely nonzero.

Now, we prove the lemma 2 in the Letter, which is stated as:

\emph{Lemma 2}. In $N$-qubit mixed states, the multipartite entanglement indicator
\begin{eqnarray}\label{s55}
\tau_{SEF}^{(1)}(\rho_N)=\mbox{min}\sum_jp_j
\frac{\sum_{l=1}^n\tau_{SEF}^{(1)}(\ket{\psi^j}^{A_l}_N)}{N}
\end{eqnarray}
is zero if and only if the quantum state is $(N/2)$-separable in the form $\rho_{A_1A_2\cdots
A_n}=\sum_{i_1,\cdots,i_n=1}^n\sum_j p_j^{\{i_1\cdots i_n\}}\rho_{Ai_1Ai_2}^j\otimes \cdots
\otimes\rho_{Ai_{k-1}Ai_{k}}^j\otimes\cdots \otimes\rho_{Ai_{n-1}Ai_{n}}^j$, which has at most
two-qubit entanglement with the superscript $\{i_1\cdots i_n\}$ being all permutations of the $N$
qubits.

\emph{Proof}. When the $N$-qubit quantum state is in the form shown in this lemma, it is easy to
obtain that all the indicators $\tau_{SEF}^{(1)}(\ket{\psi^j}^{A_l}_N)$s are zero for
$l\in\{1,2,\cdots, n\}$ according to the lemma b. Thus the multipartite entanglement indicator
$\tau_{SEF}^{(1)}(\rho_N)$ is zero, which proves the necessity of lemma 2. Next, we analyze the
sufficiency. When the indicator $\tau_{SEF}^{(1)}(\rho_N)$ is zero, we can obtain that, for each
component $\ket{\psi^{j}}_N$ in the optimal pure state decomposition of $\rho_N$, the qubit $A_l$
is entangled with at most one other qubit $A_k$ with $k\neq l$. In this case, the pure state
component is a tensor product state of at most $N/2$ two-qubit entangled states. Note that the
two-qubit tensor product state $\ket{\phi}_{A_l}\otimes\ket{\phi}_{A_k}$ is a special form of pure
state $\ket{\phi}_{A_lA_k}$. After considering all the pure state components, the $N$-qubit mixed
state can be written in a general $(N/2)$-separable form
\begin{eqnarray}\label{s56}
\rho_{A_1A_2\cdots A_n}&=&\sum_{i_1,\cdots,i_n=1}^n\sum_j p_j^{\{i_1\cdots
i_n\}}\rho_{Ai_1Ai_2}^j\otimes \cdots \nonumber\\
&&\otimes\rho_{Ai_{k-1}Ai_{k}}^j\otimes\cdots \otimes\rho_{Ai_{n-1}Ai_{n}}^j,
\end{eqnarray}
which only contains two-qubit entanglement with the superscript $\{i_1\cdots i_n\}$ being all
permutations of the $N$ qubits.  This completes the proof of lemma 2 in the Letter.

According to lemma 2, the indicator $\tau_{SEF}^{(1)}(\rho_N)$ is surely nonzero whenever an
$N$-qubit mixed state contains the genuine multiqubit entanglement.

Next, as an application case, we will analyze an $N$-qubit mixed state written as
\cite{bai08pra2-s}
\begin{eqnarray}\label{s57}
\varrho_{A_1A_2\cdots A_n}=\alpha\proj{1^{\otimes N}}+(1-\alpha)\proj{W_N},
\end{eqnarray}
where the parameter $\alpha=1/(N+1)$ and the $N$-qubit $W$ state being $\ket{W_N}=(\ket{10\cdots
0}+\ket{01\cdots 0}+\cdots +\ket{00\cdots 1})/\sqrt{N}$. In Ref. \cite{bai08pra2-s}, it was pointed
out that this quantum state is entangled but without two-qubit concurrence and multiqubit
$n$-tangles with $n=3,4,\cdots, N$, where the $n$-tangle is a kind of multipartite entanglement
measure based on the monogamy property of squared concurrence.

To reveal the critical entanglement structure, we will utilize the newly presented multipartite
entanglement indicator $\tau_{SEF}^{(1)}(\rho_N)$ in Eq. (55). Due to the quantum state
$\varrho_{A_1A_2\cdots A_n}$ being symmetric under qubit permutations, we can get the following
relation
\begin{eqnarray}\label{s58}
&&\tau_{SEF}^{(1)}(\varrho_N)=\tau_{SEF}^{(1)}(\varrho_N^{A_1})\\
&&=\mbox{min}\sum_i p_i[E_f^2(\ket{\psi^i}_{A_1|A_2\cdots A_n}) -\sum_{j\neq
1}E_f^2(\rho^i_{A_1A_j})],\nonumber
\end{eqnarray}
where the minimum runs over all the pure state decompositions of $\varrho_{A_1A_2\cdots A_n}$.
After some analysis, we obtain that the optimal decomposition is
\begin{eqnarray}\label{s59}
\varrho_{A_1A_2\cdots A_n}^{opt}=p_1\proj{\phi^1_N}+p_2\proj{\phi^2_N}
\end{eqnarray}
with the two components being $\ket{\phi^1_N}=\ket{1^{\otimes N}}$ and $\ket{\phi^2_N}=\ket{W_N}$,
which means the decomposition in Eq. (57) is optimal. Therefore, the entanglement indicator in Eq.
(58) is
\begin{eqnarray}\label{s60}
&&\tau_{SEF}^{(1)}(\varrho_N)\nonumber\\
&=&\alpha\tau_{SEF}^{(1)}(\ket{\phi^1_N})
+(1-\alpha)\tau_{SEF}^{(1)}(\ket{\phi^2_N})\nonumber\\
&=&(1-\alpha)\tau_{SEF}^{(1)}(\ket{W_N^{A_1}})\\
&=&\frac{N}{N+1}[E_f^2(C_{A_1|A_2\cdots A_n}^2)-(N-1)\cdot E_f^2(C_{A_1A_2}^2)],\nonumber
\end{eqnarray}
where we use the symmetry property of $\ket{W_N}$, and the two concurrences are $C_{A_1|A_2\cdots
A_n}^2=4(N-1)/N^2$ and $C_{A_1A_2}^2=4/N^2$. The indicator is nonzero and detects the existence of
genuine multiqubit entanglement. Since the optimal component $\ket{W_N}$ is not product under any
partition, the multipartite entanglement is the genuine $N$-qubit entanglement. In Table I, we give
the values of multiqubit entanglement indicator $\tau_{SEF}^{(1)}(\varrho_N)$ for the cases
$N=3,4,7,10,20,30$.
\begin{table}[b]
\centering
\begin{tabular}{|c|c|c|c|c|c|c|}
  \hline
  $N$ & 3 & 4 & 7 & 10 & 20 & 30\\\hline
  $\tau_{SEF}^{(1)}(\varrho_N)$ & 0.2992 & 0.2813 & 0.0992 & 0.0401 & 0.0053& 0.0015\\
  \hline
\end{tabular}
\caption{The nonzero values of indicator $\tau_{SEF}^{(1)}(\varrho_N)$ indicate the existence of
genuine $N$-qubit entanglement with $N=3,4,7,10,20,30$.}
\end{table}

\subsection{VII. The calculation of indicator $\tau_{SEF}^{(2)}(\rho_{c_1c_2r_1}^{c_1})$ in the
dynamics of cavity-reservoir systems}

In the Letter, the multipartite entanglement indicator of the system $c_1c_2r_1$ has the form
\begin{equation}\label{s61}
\tau_{SEF}^{(2)}(\rho_{c_1c_2r_1}^{c_1})=E_f^2(c_1|c_2r_1)-E_f^2(c_1c_2)-E_f^2(c_1r_1)
\end{equation}
which is used to detect the genuine tripartite entanglement in the dynamical procedure. Here, a
crucial point is to calculate the bipartite entanglement $E_f(c_1|c_2r_1)$, because the subsystem
$c_2r_1$ is not equivalent to a logic qubit and then the formula in Eq. (3) of Letter does not
work. From the Koashi-Winter formula~\cite{koashi04pra-s}, this entanglement is related to quantum
discord
\begin{eqnarray}\label{s62}
E_f(c_1|c_2r_1)&=&D(c_1|r_2)+S(c_1|r_2)\nonumber\\
&=& \mbox{min} _{\{E_k^{r_2}\}}\sum_k p_k S(c_1|E_k^{r_2}),
\end{eqnarray}
where the minimum runs over all the positive operator-valued measures (POVMs). Chen \emph{et al}
presented an effective method for calculating the quantum discord and choosing the optimal
measurement~\cite{chen11pra-s}. After some analysis, we obtain that the optimal measurement is
$\sigma_x$, and then the entanglement of formation is
\begin{equation}\label{s63}
E_f(c_1|c_2r_1)=-\eta\mbox{log}_2(\eta)-(1-\eta)\mbox{log}_2(1-\eta),
\end{equation}
where $\eta=(1-q)/2$ with $q=[1-4\beta^2\xi^2(\xi^2+\beta^2\chi^2-\beta^2\xi^2)]^{1/2}$. In Fig.2
of the main text, the squared entanglements $E_f^2(c_1|c_2r_1)$, $E_f^2(c_1c_2)$ and
$E_f^2(c_1r_1)$ are plotted as functions of the time evolution $\kappa t$ and the initial amplitude
$\alpha$, which characterize the bipartite entanglement distribution in the multipartite system. In
particular, the indicator $\tau_{SEF}^{(2)}(\rho_{c_1c_2r_1}^{c_1})$ in the figure detects the
genuine tripartite entanglement in the dynamical procedure.

\subsection{VIII. The indicator $\tau_{SEF}^{(2)}$ not being monotonic under local operations
and classical communication}

The entanglement indicator $\tau_{SEF}^{(2)}$ in Eq. (16) of Letter is used to detect the
multipartite entanglement not stored in two-qubit pairs. In the following, we will prove that this
indicator is not an entanglement monotone under local operations and classical communication
(LOCC).

As a counter-example, we consider the tripartite entanglement indicator
$\tau_{SEF}^{(2)}(\rho_{c_1c_2r_2}^{c_1})$, in which $\rho_{c_1c_2r_2}$ is the reduced state of
$\ket{\Phi_t}_{c_1c_2r_1r_2}$ in Eq. (19) of the main text. When the initial state parameter is
chosen as $\alpha=9/10$ and the time evolution is $\kappa t=0.9$, the value of the indicator is
\begin{eqnarray}\label{s64}
\tau_{SEF}^{(2)}(\rho_{c_1c_2r_2}^{c_1})
&=&E_f^2(c_1|c_2r_2)-E_f^2(c_1c_2)-E_f^2(c_1r_2)\nonumber\\
&=&0.0925.
\end{eqnarray}
It is known that any local protocol can be decomposed into a sequence of two-outcome POVMs
involving only one party~\cite{dur00pra-s}. For the quantum state $\rho_{c_1c_2r_2}$, we consider a
local POVM performed on the subsystem $c_1$, which has the form~\cite{dur00pra-s}
\begin{eqnarray}\label{s65}
M_1=\left(
      \begin{array}{cc}
        a & 0 \\
        0 & b \\
      \end{array}
    \right),
M_2=\left(
      \begin{array}{cc}
        \sqrt{1-a^2} & 0 \\
        0 & \sqrt{1-b^2} \\
      \end{array}
    \right),
\end{eqnarray}
and satisfies the relation $M_1^\dagger M_1+M_2^\dagger M_2=I$. Here we choose the POVM parameters
to be $a=4/5$ and $b=3/7$, respectively. After the local operation, the average entanglement
indicator is
\begin{eqnarray}\label{s66}
&&<\tau_{SEF}^{(2)}(\rho_{c_1c_2r_2}^{c_1})>
=p_1\tau_{SEF}^{(2)}(\rho_{1})+p_2\tau_{SEF}^{(2)}(\rho_{2})\nonumber\\
&=&0.6047\times 0.0157+0.3953\times 0.2376\nonumber\\
&=&0.1034,
\end{eqnarray}
where $\rho_i=M_i\rho_{c_1c_2r_2}M_i^\dagger/p_i$ with
$p_i=\mbox{tr}[M_i\rho_{c_1c_2r_2}M_i^\dagger]$ and $i\in\{1,2\}$, respectively. By a direct
comparison, we have
\begin{eqnarray}\label{s67}
&&<\tau_{SEF}^{(2)}(\rho_{c_1c_2r_2}^{c_1})>-\tau_{SEF}^{(2)}(\rho_{c_1c_2r_2}^{c_1})\nonumber\\
&&=0.1034-0.0925\nonumber\\
&&=0.0109,
\end{eqnarray}
which means that the indicator $\tau_{SEF}^{(2)}(\rho_{c_1c_2r_2}^{c_1})$ can increase under the
LOCC. Therefore, we obtain that the multipartite entanglement indicator $\tau_{SEF}^{(2)}$ in Eq.
(16) of Letter is not an entanglement monotone.

Finally, we want to point out that $\tau_{SEF}^{(2)}$ is still an effective indicator for
multipartite entanglement. The situation is similar to that of residual tangle for the squared
concurrence which is also increasing under the LOCC~\cite{byw07pra-s} but can be served as a
quantifier for multipartite entanglement~\cite{afov08rmp-s}.


\begin{thebibliography}{99}
\bibitem{note1}             Note that the Wootter's formula of Eq. (3) cannot be directly applied
                            to the entanglement of formation $E_f(\rho_{A_1|A_2\cdots A_n})$ because
                            the subsystem $A_2\cdots A_n$ is not a logic qubit in general.
\end{thebibliography}
\end{document}